\documentclass[12pt]{article} 
\usepackage[tablesfirst,notablist,nomarkers]{endfloat} 

\usepackage{ol2}
\usepackage[draft]{hyperref}
\usepackage{amsmath}

\begin{document}
%
%
\title{Critical coupling in a system of two coupled negative index medium layers}
\author{Subimal Deb and S. Dutta Gupta$^*$}
\address{School of Physics, University of Hyderabad, Hyderabad 500046, India\\
$^*$Corresponding author: sdghyderabad@gmail.com}
\begin{abstract}
We study critical coupling (CC) in a system of two coupled negative index medium (NIM) layers leading to near-total suppression of both reflection and transmission at specified frequencies. The tunability of the CC frequency is demonstrated by varying the angle of incidence retaining the full causal response for the NIM materials.
\end{abstract}
\maketitle 
%
%
In recent years there has been a surge of activities on layered media \cite{yeh, sdg1998}. The research is directed not only to well-studied guided/surface modes but also to novel applications ranging from sensors to slow light, from strong coupling regime of cavity QED to near-perfect absorption \cite{debackere, sdgpramana,  tischler2005, MangaRao2004}. Critical coupling (CC) leading to suppression of both reflection and transmission (and hence near-perfect absorption) at a given frequency was reported by Tischler {\em et al} \cite{tischler2006}. They used a 5 nm thick layer of highly absorbing polymer, separated from a dielectric Bragg reflector (DBR) by a spacer layer under normal incidence. The polymer layer was later replaced by a metal colloid film to show the possibility of critical coupling simultaneously at two distinct frequencies \cite{sdgol2007}. Subsequently these studies were extended to oblique incidence to show the feasibility of critical coupling both for TE and TM polarized light \cite{sdgjopa}. The choice and the tunability of the CC frequency are limited by the spectral properties of the absorbers and the DBR. For example, for increasing angle of incidence the unaffected absorber frequency can be outside the DBR stop-gap due to the shift of the latter and the CC effect dies. In this letter we show that both these limitations can be overcome by using metamaterials or negative index materials (NIMs), albeit in a non-standard spectral domain. The possibility of having both the permittivity and permeability negative was first pointed out by Veselago \cite{veselago}, who listed out some of the unusual properties of these metamaterials not occurring in nature. Intensive research started off from the prediction of perfect lensing and the experimental realization of these materials \cite{pendry2000, shelby2001}. Now the list of possible applications is truly significant and it is evergrowing. Very recently yet another application of a slab of metamaterial was pointed out by Bloemer {\em et al} \cite{bloemer2005}. Usually in the context of metamaterials, one considers the frequency domain where the permittivity and permeability are negative. They considered a different spectral domain between the magnetic and the electric plasma frequencies and showed that the NIM slab can exhibit a stop-band (like in 1D photonic band gap structure). They discussed in detail the similarity and differences between the two structures. In this letter we show that the stop band of the NIM material can be exploited effectively to achieve frequency tunable critical coupling. In particular, we study a system of two NIM layers (with two distinct sets of parameters) separated by a dielectric layer. The parameters are chosen such that the stop band of one falls inside the stop band of the other. The critical coupling is mediated by the Fabry-Perot (FP) resonances of the dielectric layer sandwiched  between the NIM layers. The cavity formation is allowed since the real part of the refractive index of both the NIMs in the said frequency domain is close to zero and they act like barriers. We demonstrate a great flexibility with respect to the choice of the CC frequency, even to the extent of simultaneous CC at two well-separated frequencies.
\par 
We first recall some of the important features of a NIM slab in the context of omnidirectional reflection. Let the frequency dependence of the permittivity and permeability of the NIM be given by \cite{shelby2001}
\begin{eqnarray}\nonumber
\epsilon(f)&=& 1-\frac{f_{ep}^2-f_{eo}^2}{f^2-f_{eo}^2+i\gamma f}\\
&& \label{eq:epsmu}\\ 
\mu(f)     &=& 1-\frac{f_{mp}^2-f_{mo}^2}{f^2-f_{mo}^2+i\gamma f}\nonumber
\end{eqnarray}
where $f_{ep}$ ($f_{mp}$) is the electric (magnetic) plasma frequency, $f_{eo}$ ($f_{mo}$) is the electric (magnetic) resonance frequency and $\gamma$ is the decay rate. Experimental 
\begin{figure}\centering
\includegraphics[width=8.5cm]{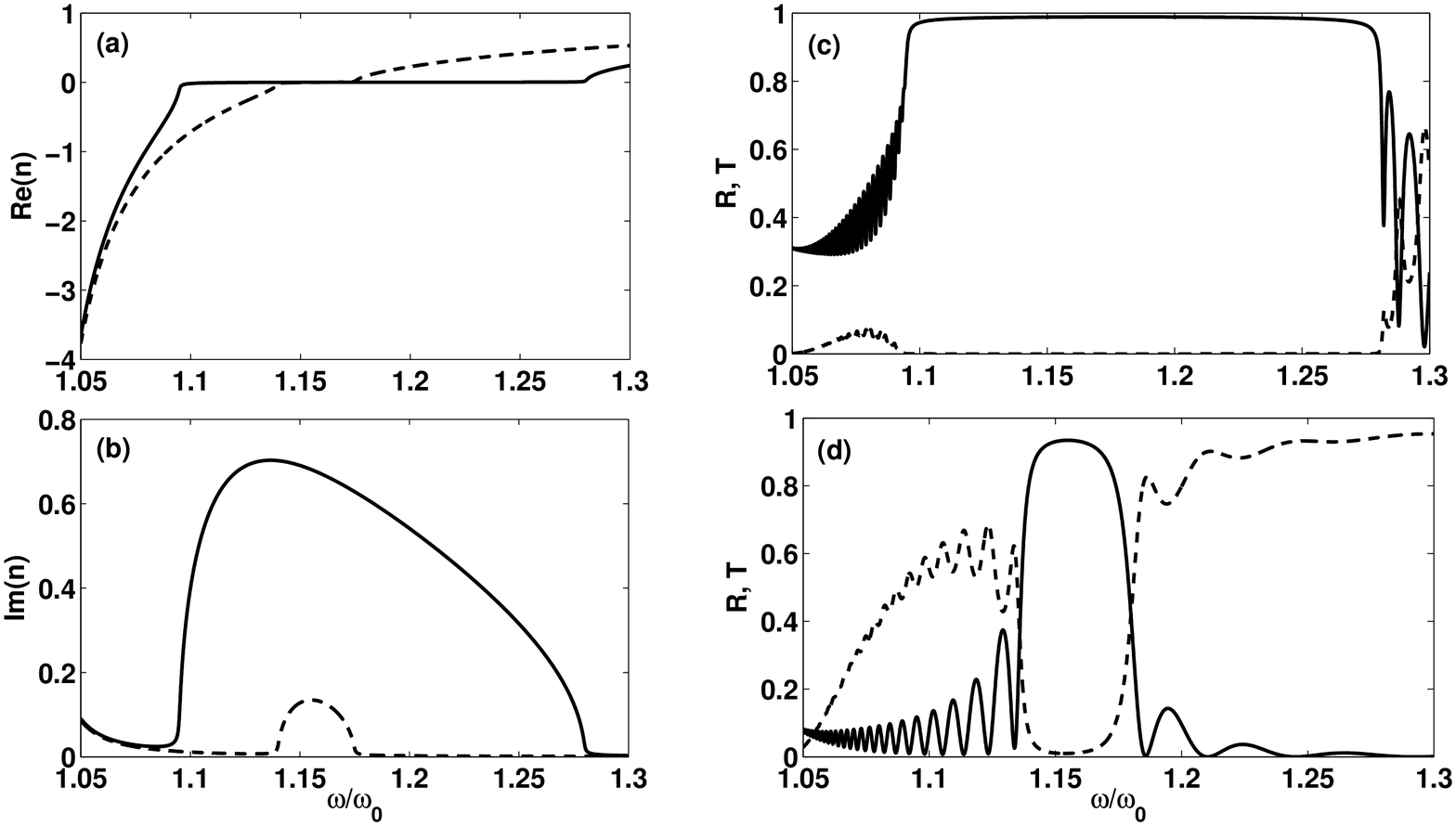}
\caption{Left: (a) Real and (b) imaginary parts of the refractive indices of NIM$_t$ (dashed) and NIM$_b$ (solid). Right: The reflection (solid) and transmission (dashed) profiles of a single slab for (c) NIM$_b$, $d=5$ and (d) NIM$_t$, $d=3$.}\label{fig:nims}
\end{figure}
demonstration of NIMs \cite{shelby2001} for TE polarization had been made using a 2D anisotropic structure. We assume our NIM material to be 3D and isotropic with macroscopic properties given by Eq. (\ref{eq:epsmu}). In the context of a slab of such a material, Bloemer {\it et al} \cite{bloemer2005} demonstrated omnidirectional reflectivity in the frequency band between the electric and magnetic plasma frequencies. Within this band the real part of the refractive index is zero. With increasing angle of incidence the stop band was shown to widen with the higher frequency edge moving to even higher frequencies and the lower frequency edge practically insensitive. We use two NIM slabs described by Eq. (\ref{eq:epsmu}) albeit with distinct set of parameters such that the stop band of one falls inside the stop band of the other. In order to deal with dimensionless quantities (like in \cite{bloemer2005}), all the frequencies will be scaled in units of $f_0=c/\lambda_0=10$ GHz and the lengths in units of $\lambda_0$. The parameters for one of the NIMs (labeled as NIM$_b$) are chosen from the experiment of Shelby \cite{shelby2001}, i.e., $f_{ep}=1.28$, $f_{mp}=1.095$, $f_{mo}=1.005$, $f_{eo}=1.03$ and $\gamma=0.001$. For the other NIM (labeled by NIM$_t$) the parameters are the same, except for $f_{ep}=1.175$, $f_{mp}=1.14$. The real and imaginary parts of the two NIMs are shown in Figs. \ref{fig:nims}(a) and \ref{fig:nims}(b), respectively. The reflectivity and transmittivity profiles for slabs of NIMs with width $d$ (calculated using the characteristic matrix method \cite{bornwolf, sdg1998} ) for normal incidence are shown in Figs. \ref{fig:nims}(c) and \ref{fig:nims}(d). The occurrence of the stop band of NIM$_t$ inside that of NIM$_b$ can easily be discerned from these plots. The feature that is crucial for our applications is the near-zero real part of the refractive index and broadband absorption over the stop bands. This is in contrast to the use of narrow band absorbers in earlier realizations of CC \cite{tischler2006, sdgol2007, sdgjopa}.
\begin{figure}\centering
\includegraphics[width=5cm]{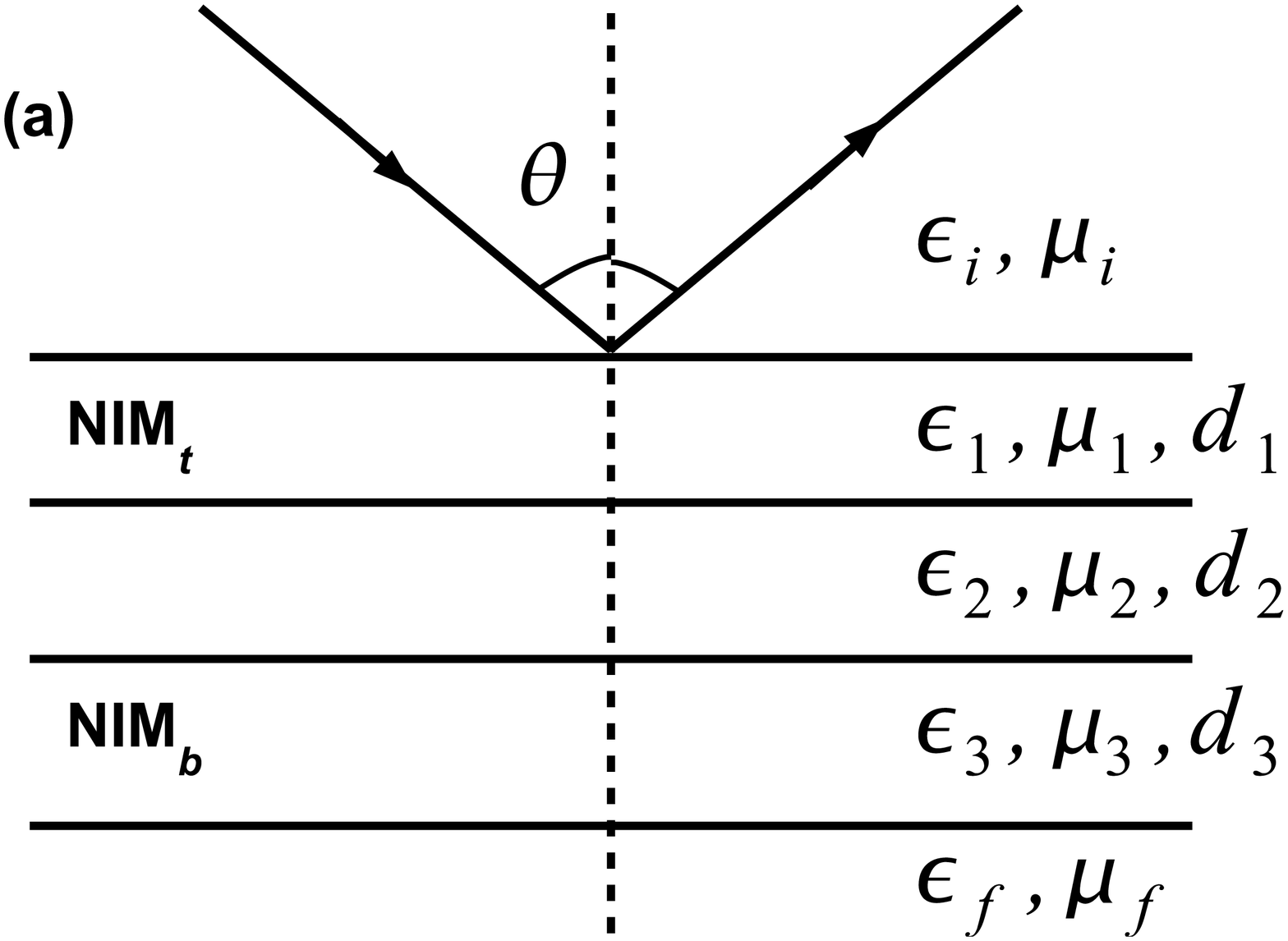}\\
\includegraphics[width=8.7cm]{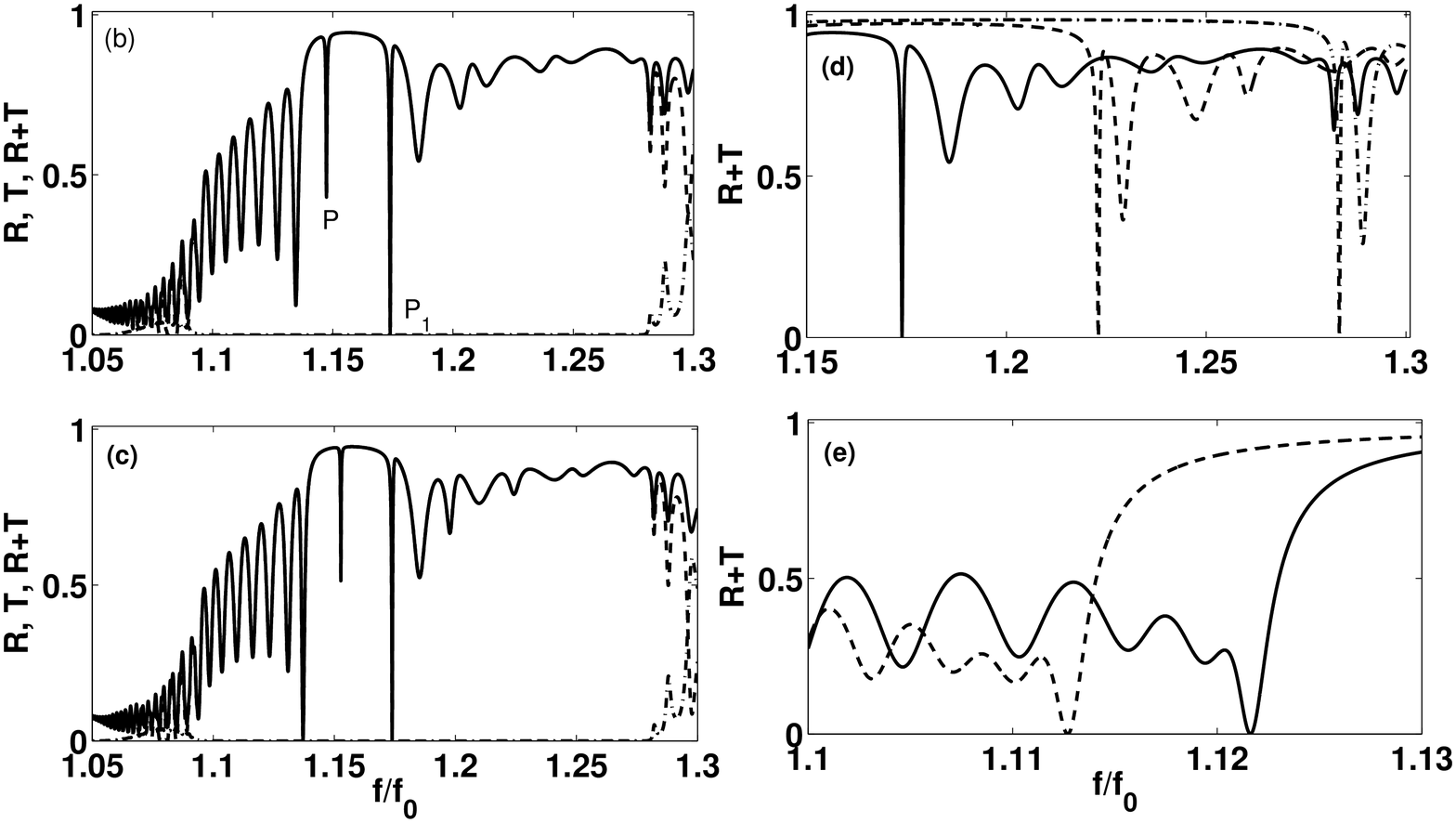}
\caption{(a)Schematic of the layered structure. 
$R$(dashed), $T$(dash-dot) and $R+T$(solid) showing CC at normal incidence (b) near the right band edge. 
(c) simultaneously at two frequencies near the band edges. 
$R+T$ showing CC at discrete $\theta$ (d) near the right band edge at (from left to right, $\theta=0^\circ$, $18.5^\circ$, $29.4^\circ$) 
(e) near the left band edge at (from right to left $\theta=18^\circ$, $26.5^\circ$). (b), (d) and (e) are for $d_2=7.042$. (c) is for $d_2=9.664$.}\label{fig:thetaopt}
\end{figure}
\par 
Our structure (shown in Fig. \ref{fig:thetaopt}(a)) consists of two NIM layers (top NIM$_t$ and bottom NIM$_b$ with thickness $d_1$ and $d_3$, respectively), separated by a dielectric slab of dielectric constant $\epsilon_2$, permeability $\mu_2$ and thickness $d_2$. The objective is to achieve critical coupling within the smaller stop band of NIM$_t$. The structure is embedded in air with $\epsilon_i=\epsilon_f=1$, $\mu_i=\mu_f=1$. Henceforth, we  consider the incidence of TE-polarized plane wave at an angle $\theta$, though similar results  can be obtained for TM-polarized light also. Note that the spacer layer, due to its higher refractive index compared to its neighbors forms a Fabry-Perot cavity with its own Airy resonances. These modes are affected by the high losses in the NIMs and associated phase change at the interfaces. The signature of these modes are imprinted on the total scattering $R+T$, where $R$ and $T$ are the intensity reflection and transmission of the structure.
\par 
We now demonstrate CC at single and simultaneously at two frequencies by a suitable choice of the spacer layer thickness. In Fig. \ref{fig:thetaopt}(b) we show CC near the right band edge (at $f=1.174$) for normal incidence for $d_2\sim7$. As mentioned earlier, the dips in the reflectivity profile at the points P and P$_1$ are due to FP resonances of the dielectric spacer layer. This can be checked in two ways. First, they broaden upon introduction of absorption in the dielectric spacer material (curves not shown). Secondly, the feature at P$_1$ can be recovered again by increasing $d_2$ by $(\Delta d_2)_1 = 1/(2n_2 f_1)$ (in scaled units), which adds a phase of $2\pi$ to the reflected wave. Here $f_1$ refers to the CC frequency at P$_1$.  This CC frequency has a small tunability over a small frequency domain if the angle of incidence is varied within $\pm 2^0$. CC can also be achieved at another frequency near the left edge for a suitable choice of $d_2$. Note that combining the FP resonances with the band edge features makes it easier for achieving CC. It is clear from above discussions that CC near the right band edge can be retained, say, at frequency $f_1$ by increasing/decreasing $d_2$ in steps of $(\Delta d_2)_1$. This change in $d_2$ alters the phase at the other frequencies. Thus by changing $d_2$, CC can be achieved simultaneously at another frequency $f_2$ when $m/(2n_2 f_1) = p/(2n_2 f_2)$ ( $m$ and $p$ are integers). This ensures phase change in multiples of $\pi$ for a single pass through the spacer layer at both the frequencies $f_1$ and $f_2$. Simultaneous CC near the two band edges was achieved in this way (Fig. \ref{fig:thetaopt}(c)). 
\par 
Recalling that a change in the angle of incidence  causes a change in the width of the stop band of the NIM slab, we show in Figs. \ref{fig:thetaopt}(d) and \ref{fig:thetaopt}(e) that CC can be achieved at different frequencies near the edges of the widened band. We show this for two angles of incidence near the left (for $\theta=18^\circ, 26.5^\circ$) and right (for $\theta=18.5^\circ, 29.4^\circ$) band edges. Figs. \ref{fig:thetaopt}(d) and \ref{fig:thetaopt}(e) thus demonstrate the remarkable flexibility of changing the CC frequency by a mere tilt of the structure with respect to incident light. It is to be noted that CC at oblique incidence, as for normal incidence, originates again from the interplay of the FP resonances of the spacer layer with the stop band features of the NIM.
\par 
\begin{figure}\centering
\includegraphics[width=8.3cm]{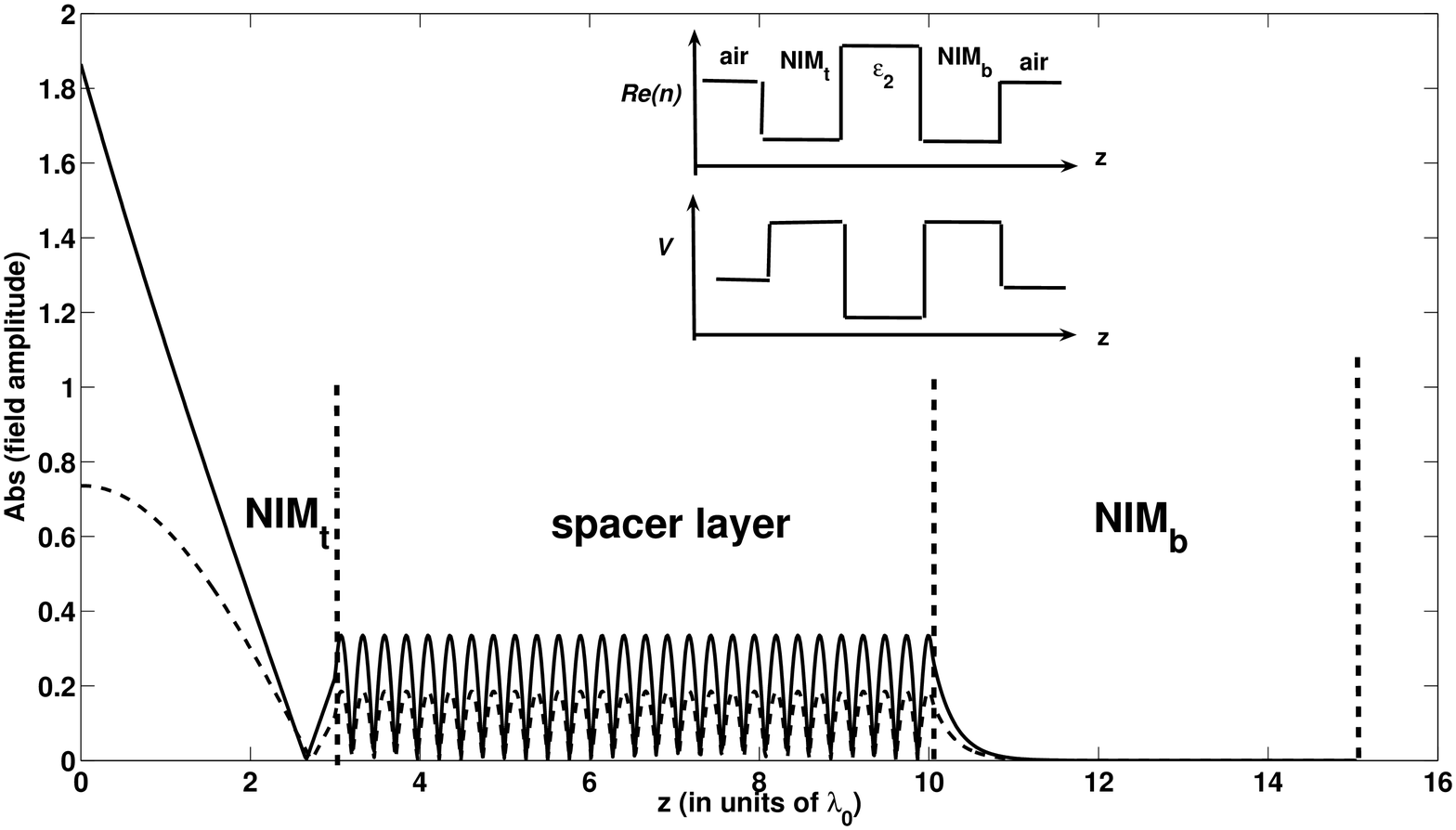}
\caption{Field distribution at ($f=1.174$, solid curve) and away ($f=1.18$, dashed curve) from the CC frequency for the parameters in Fig. \ref{fig:thetaopt}(b). The vertical dashed lines mark the interfaces of the layered structure. (Inset) Schematic of the refractive index distribution and its equvalent picture of a quantum mechanical potential.}\label{fig:distribz}
\end{figure}
The fact that almost  all the incident energy is dumped in the structure at a CC frequency can be illustrated by the field distribution inside the layered structure (Fig. \ref{fig:distribz}). In Fig. \ref{fig:distribz} we have plotted the magnitude of the electric field amplitude at and slightly away from the CC frequency (see the dip marked by P$_1$ in Fig. \ref{fig:thetaopt}(b)). The field amplitudes are normalized with respect to the incident field amplitude at the CC frequency. A drop in the energy retained in the structure for a non-critical frequency, highlights the near-perfect absorption at the CC frequency. In order to gain some insight into the field distributions and the resonances of the structure we look at a simplified model (neglecting losses) of an equivalent quantum mechanical scattering problem. The equivalence of one-dimensional optical and quantum mechanical problems is now well understood \cite{kay, sdgopex}. The refractive index profile and the corresponding potential (ignoring losses) are shown in the inset of Fig. \ref{fig:distribz}. In the frequency range where both the NIMs have null (real part) refractive index, the NIM layers act like barriers enclosing the well formed by the high index dielectric layer. One thus has the possibility of exciting the bound states of this well (these are the FP modes). The real scenario is much more complicated, because of the losses in the NIMs. For example, the barrier represented by the bottom NIM is practically impenetrable because of the very high losses. Thus there is practically no transmission through the structure. The `bound' states of the well dig holes in the otherwise total reflection resulting in the critical coupling.
\par 
Finally, let us compare the present scheme of achieving CC with the earlier methods making use of the narrow absorption resonances of polymer or silver colloid films. The substrate used in the earlier structures was a DBR, while the current one uses a NIM layer for broadband reflection. The advantages of using a NIM was already pointed out by Bloemer {\it et al} \cite{bloemer2005}, who showed that there is a significant broadening  of the bandgap (with `floating' right and `static' left edges) under oblique incidence. In case of a DBR the stop band shifts to the right with not so significant broadening. In the earlier scheme, the role of the spacer layer was mainly to control the phases so that there is destructive interference in the medium of incidence leading to the suppression of reflection. In the present scheme, the role of the spacer layer is two fold, namely it adjusts the phases. At the same time its resonances (there are multitudes of them) hold the key to the CC. The easy control of these resonances leads to the overall flexibility of the CC phenomenon of the total structure. Note that earlier schemes suffered from the rigidity that CC could be achieved only at or near the absorption frequency of the polymer or the colloid film. Moreover, these frequencies need to be within the stop band of the DBR.
\par 
We have shown critical coupling with TE polarized light in a layered structure with two NIM slabs coupled by a dielectric spacer layer. Both the NIM slabs are modeled by a full causal response. CC at a single and a pair of frequencies have been demonstrated. Continuous tunability for small angles of incidence and discrete tunability at larger angles have been demonstrated. Tuning the system at different frequencies is achieved by a simple change of the angle of incidence. These flexible attributes are shown to follow from the omnidirectional stop band features of the NIM slabs. CC should be useful for devices where near-perfect absorption of incident light is desired at one or more frequencies.
\par
The authors are thankful to the Department of Science and Technology, Government of India, for support. One of the authors (SDG) is also thankful to the Nano Initiative Program of University of Hyderabad for partial support.
\end{document}